\documentclass[usegraphicx, usenatbib]{mn2e}
\usepackage{times}
\usepackage{amsmath}

\def\kms{\ifmmode\,{\rm km}\,{\rm s}^{-1}\else km$\,$s$^{-1}$\fi}

\def\msun{\,{\rm M_{\sun}}}

\title[Binarity in Carbon-Enhanced Metal-Poor stars]{Binarity in Carbon-Enhanced Metal-Poor stars\thanks{Based on observations obtained with the Hobby-Eberly Telescope, which is a joint project of the University of Texas at Austin, the Pennsylvania State University, Stanford University, Ludwig-Maximilians-Universit\"{a}t M\"{u}nchen, and Georg-August-Universit\"{a}t G\"{o}ttingen.}} 
\author[E. Starkenburg et al.]{Else Starkenburg$^{1}$\thanks{email: else@uvic.ca}\thanks{CIFAR Global Scholar}, Matthew D. Shetrone$^{2}$, Alan W. McConnachie$^{3}$ and Kim A. Venn$^{1}$\\
$^{1}$ Dept. of Physics and Astronomy, University of Victoria, P.O. Box 3055, STN CSC, Victoria BC V8W 3P6, Canada \\
$^{2}$ McDonald Observatory, The University of Texas at Austin, 1 University Station, C1400, Austin, TX 78712-0259, USA \\
$^{3}$ NRC Herzberg Institute of Astrophysics, 5071 West Saanich Road, Victoria, BC, V9E 2E7, Canada} 
\begin{document}

\maketitle

\begin{abstract}
A substantial fraction of the lowest metallicity stars show very high enhancements in carbon. It is debated whether these enhancements reflect the stars' birth composition, or if their atmospheres were subsequently polluted, most likely by accretion from an AGB binary companion. Here we investigate and compare the binary properties of three carbon-enhanced sub-classes: The metal-poor CEMP-s stars that are additionally enhanced in barium; the higher metallicity (sg)CH- and Ba II stars also enhanced in barium; and the metal-poor CEMP-no stars, not enhanced in barium. Through comparison with simulations, we demonstrate that all barium-enhanced populations are best represented by a $\sim$100\% binary fraction with a shorter period distribution of at maximum $\sim$20,000 days. This result greatly strengthens the hypothesis that a similar binary mass transfer origin is responsible for their chemical patterns. For the CEMP-no group we present new radial velocity data from the Hobby-Eberly Telescope for 15 stars to supplement the scarce literature data. Two of these stars show indisputable signatures of binarity. The complete CEMP-no dataset is clearly inconsistent with the binary properties of the CEMP-s class, thereby strongly indicating a different physical origin of their carbon enhancements. The CEMP-no binary fraction is still poorly constrained, but the population resembles more the binary properties in the Solar Neighbourhood.
\end{abstract}


\section{Introduction}

The lowest metallicity stars that still exist today probably carry the
imprint of very few supernovae. As such, they represent our best
observational approach to understanding of the First Stars. The number of metal-poor stars in the Galactic halo 
with abundance determinations has bloomed recently; 
now over 150 stars with [Fe/H] $< -3$ have been examined in high-resolution studies \citep[see for some recent overviews and results:][]{Aoki13, Yong13a, Cohen13, Spite13, Placco14}. In addition to stars with a “normal” chemical composition (those continuing the well-defined trends of 
elemental abundances at higher metallicities), various chemically peculiar stars are
found.  The intriguing
question therefore is if these chemical anomalies tell us something
about the very first stages of star formation? 
 
Probably the most significant chemical sub-group is that of
carbon-enhanced metal-poor (CEMP) stars. The fraction of extremely
metal-poor stars which are carbon-enhanced 
is as high as 32-39\% for stars with [Fe/H]$<$-3 \citep{Yong13b,Aoki13,Lee13}, which then decreases to 9-21\% for [Fe/H]$<$-2.0
\citep[e.g.,][]{Norris97a,Rossi99,Christlieb03,Marsteller05,Lucatello06,Cohen05,Frebel06,Carollo12,Lee13}\footnote{We
  caution that various definitions of carbon-enrichment are used
  between different authors. Generally, a limit of [C/Fe]=+1.0 or
  [C/Fe]=+0.7 is used, although some authors prefer a limit that is
  dependent on the luminosity of the star to take into account
  internal mixing of carbon on the red giant branch
  \citep[e.g.][]{Aoki07}.}. The fraction of CEMP stars is not
just dependent on the metallicity, but also seems to vary with the
distance above the Galactic plane \citep{Frebel06,Carollo12} and between the inner and outer halo component \citep{Carollo12,Carollo14}. Measurements of [C/Fe] for extremely metal-poor stars range all the way up to 
[C/Fe]$\sim+4.0$. 

A substantial fraction of CEMP stars also show an overabundance in heavy elements, these are generally called  
``CEMP-s'', ``CEMP-r'', or ``CEMP-r/s'' depending on the exact
abundance and ratio of s-process and r-process
elements. \citet{Beers05} have developed the following nomenclature 
that we will follow in this paper (note that the different classes as defined here are not necessarily mutually exclusive):
\begin{itemize}
\item{CEMP-r:  [C/Fe] $>$ +1.0 and [Eu/Fe] $>$ +1.0}
\item{CEMP-s:  [C/Fe] $>$ +1.0, [Ba/Fe] $>$ +1.0, and [Ba/Eu] $>$ +0.5}
\item{CEMP-r/s: [C/Fe] $>$ +1.0 and 0.0 $<$ [Ba/Eu] $<$ +0.5} 
\item{CEMP-no: [C/Fe] $>$ +1.0 and [Ba/Fe] $<$ 0}
\end{itemize} 

The large class of CEMP-s stars are thought to obtain their
overabundant carbon and s-process elements from a companion star that has gone through the
AGB phase and deposited large amounts of newly formed carbon and
s-process material on its neighbour. Strong evidence in favour of this scenario was found from repeated radial velocity 
measurements. The fraction of stars that show significant velocity variability in the CEMP-s class is sufficiently high to 
comfortably support the claim that all such stars might indeed be in binary systems 
\citep[][and references therein]{Lucatello05}.

Based on their abundance patterns, it has been suggested that the CEMP-r/s class has the same binary origin as the CEMP-s stars \citep[e.g.,][]{Masseron10,Allen12}, although this might mean that a different neutron-capture process with features in between the r- and s-process will need to be invoked \citep[][Herwig et al., in preparation]{Lugaro12} to provide for its particular chemical signatures. An alternative explanation is that CEMP-r/s stars originate in regions already enriched in r-process elements and are subsequently enriched in s-rich material by a companion \citep[e.g.,][]{Bisterzo12}. 

CEMP-r stars are very rare, but \citet{Hansen11} studied one CEMP-r star in their careful radial velocity monitoring program of r-process-enhanced stars. This star, CS~22892--052, shows no sign of binarity.

The origin of the CEMP-no class is debated \citep[e.g.][]{Ryan05,Masseron10,Norris13b}. The absence of the signature s-process overabundance, which is thought to be produced in AGB stars just as the overabundant carbon, gives reason to believe these stars might not obtain their peculiar abundance pattern due to mass transfer in binary systems. Under the premise that CEMP-no stars are \textit{not}
in binary systems, another explanation has to be offered for the
overabundance of carbon and other light elements in these stars. One
proposed origin for their chemical pattern is that these stars are truly second generation stars and formed from gas 
clouds already imprinted with a large overabundance of carbon and other light elements 
by the First Stars \citep[e.g.][]{Bromm03,Norris13b,Gilmore13}. The fact that almost all of the stars with [Fe/H]$\le$-4.0 are of the CEMP-no class seems to favor such an explanation. In particular, four out of the five stars known with [Fe/H]$<$-4.5 seem to be consistent with the CEMP-no class \citep[][although in some cases only an upper limit for barium could be derived]{Christlieb04,Frebel05,Aoki06,Norris07,Keller14}. Note also the exception from \citet{Caffau11}. 

On the other hand, it is yet insufficiently understood if carbon could be transferred from an AGB
companion without s-process elements
\citep[e.g.,][]{Suda04}. Mass-transfer mechanisms that would transfer
carbon -- but no or few s-process elements -- are theoretically
expected from massive AGB stars with hot dredge-up, terminating the
AGB process before the star had time to develop s-process elements
\citep{Herwig04}, or some rotating AGB companions
\citep{Herwig03,Siess04}. However, this result is dependent on the parameters adopted, as shown by \citet{Piersanti13}. \citet{Komiya07} argue that relative high-mass AGB stars could be the companions of CEMP-no stars as they produce less s-process elements. But a problem with this scenario is that these stars would produce a lot of nitrogen, a signature that not all CEMP-no stars share \citep[see for instance][]{Ito13,Norris13b}. It has also been suggested that in very low-metallicity AGB-stars with very high neutron-to-Fe-peak-element seed ratios, the s-process runs to completion and a large overabundance of Pb is produced instead of Ba \citep{Busso99,Cohen06}. Because Pb absorption lines are very weak and the strongest line in the optical overlaps with the CH-feature, this hypothesis is difficult to test, especially in C-rich stars. A robust upper limit could nonetheless be given for the brightest CEMP-no star, BD +44-493, which did not show the predicted overabundance in Pb \citep{Ito13}. 

In analogy with the work on CEMP-s stars by \citet{Lucatello05} and others, we might be able to settle this debate using radial velocity monitoring. If CEMP-no stars are also products of binary evolution, this would show itself in radial velocity variations of the stars. From such an exercise, \citet{Norris13b} conclude that there is little support for a binary origin for CEMP-no stars, unlike with the CEMP-s stars. But,
as an overview of the 
available literature data such as presented in Table 5 of \citet{Norris13b} makes clear, there is a lack 
of systematic radial velocity studies with sufficient accuracy and
cadence to carry out a conclusive quantitative study. For 43\% of the stars there is only one radial velocity measurement available, making it impossible to tell whether they are part of a
binary system. Most other stars have less than 5 measurements published in the literature. The two stars that have been most thoroughly researched, BD +44-493 and CS~22957-027, do show evidence for velocity
variations, but these variations are comparable to the observational uncertainties in the case of
BD +44-493. Radial velocity data for eight more stars has in the
meantime been added \citep[][also Andersen et al., in preparation]{Hansen13}. They find that two out of their sample of eight CEMP-no stars are in binary systems. Due to these small numbers, it is not at all 
clear if CEMP-no stars have binary companions, and the presence of a companion may well 
influence their evolution. 

This current scarcity of data 
severely limits our understanding of the very first epochs of star
formation. In this work, we present additional radial velocity measurements for
 15 CEMP-no stars. Additionally, we homogeneously analyze and model the binary fraction and period distribution of binaries in the CEMP-no class, the CEMP-s class and the --  much more metal-rich -- CH-, sgCH- and Ba II-stars. Based on the comparative binary properties of each of these classes, we then go on to discuss their nature. 

In Section \ref{sec:data} we present the new data from this work. In Section \ref{sec:radvel} we use these data for the CEMP-no stars to analyze velocity variations and constrain their binary properties. Section \ref{sec:sims} is devoted to comparison with simulations for the CEMP-no, CEMP-s and CH-, sgCH- and Ba II-stars. This analysis leads to various conclusions and hypotheses for the nature of these stars, as discussed in Section \ref{sec:disc}. 

\begin{table*}
\footnotesize{
\begin{tabular}{|l|c|c|r|r|r|r|r|r|c|c|c|c|}
\hline
{Star}  &  Vmag & {T$_{\textnormal{eff}}$}  &{logg} &{[Fe/H]} & {[C/Fe]}& {[Ba/Fe]} & {\# Vr} & {Source} & $p$($\chi^{2}|f$) & $p$($\chi^{2}|f$)& $p$($\chi^{2}|f$) & $p$($\chi^{2}|f$)\\
       &   & & & & & & lit. & & lit. & this work & both & both\\
       &    & (K)  &  & & & & meas. & &1$\sigma$ &1$\sigma$ &1$\sigma$ & 3$\sigma$\\
\hline
53327-2044-515\_d                   &   15.1 & 5703 & 4.68      & $-$4.00         & +1.13          &$<$ +0.34     & 3    & 1, 13                      &  0.62 & 0.11 & 0.00 & 0.53\\ 
53327-2044-515\_g                   &   15.1 & 5703 & 3.36      & $-$4.09         & +1.57          &$<-$0.04    & 3    & 1, 13                      &  -- & -- & -- & -- \\ 
BD~+44-493                           &    9.1 & 5510 & 3.70      & $-$3.68         & +1.31          &  $-$0.59   & 28   & 2, 14                      &  0.02 & 0.92 & 0.00 & 1.00\\
BS~16929-005                        &   13.6 & 5229 & 2.61      & $-$3.34         & +0.99          &  $-$0.41   & 3    & 1, 7, 8                    &  0.06 & 0.82 & 0.00 & 0.74\\
CS~22878-027                        &   14.8 & 6319 & 4.41      & $-$2.51         & +0.86          & $<-$0.75   & 2    & 1, 8                       &  0.00 & 0.58 & 0.04 & 0.98\\ 
CS~22949-037                        &   14.4 & 4958 & 1.84      & $-$3.97         & +1.06          &  $-$0.52   & 10    & 1, 4, 9-11, 16-19          &  0.21 & 0.72 & 0.32 & 1.00\\
CS~22957-027                        &   13.6 & 5170 & 2.45      & $-$3.19         & +2.27          &  $-$0.80   & 15   & 1, 5, 12, 20, 23               &  0.00 & 0.95 & 0.00 & 0.00\\
CS~29502-092                        &   11.9 & 5074 & 2.21      & $-$2.99         & +0.96          &  $-$1.20   & 3    & 1, 8                       &  0.00 & 0.92 & 0.00 & 0.19\\
HE~1150$-$0428                      &   14.9 & 5208 & 2.54      &$-$3.47         & +2.37          &  $-$0.48   & 2    & 1, 5, 23                       &  0.00  & 0.00 & 0.00  & 0.00\\
HE~1300+0157                        &   14.1 & 5529 & 3.25      & $-$3.75         & +1.31          & $<-$0.85   & 4    & 1, 4, 6, 15                &  0.65 & 0.50 & 0.72 & 1.00\\
HE~1506$-$0113                      &   14.8 & 5016 & 2.01      & $-$3.54         & +1.47          &  $-$0.80   & 4    & 1, 13                      &  0.30 & 0.00 & 0.00   & 0.00\\
Segue~1-7                           &   17.7 & 4960 & 1.90      & $-$3.52         & +2.30          & $<-$0.96   & 1    & 3                          &  -- & 0.83 & 0.58   & 0.97\\
\hline
SDSS~J1422+0031                     &   16.3 & 5200 & 2.2       & $-$3.03         & +1.70          &  $-$1.18   & 2    & 21, 22                     &  0.01 & -- & 0.00   & 0.01 \\
SDSS~J1613+5309                     &   16.4 & 5350 & 2.1       & $-$3.33         & +2.09          &  +0.03      & 2    & 21, 22                     &  0.40 & 0.46 & 0.74 & 0.99 \\
SDSS~J1746+2455                     &   15.7 & 5350 & 2.6       & $-$3.17         & +1.24          &  +0.24      & 2    & 21, 22                     &  0.97 & 0.58 & 0.43 & 0.98\\
SDSS~J2206-0925                     &   14.9 & 5100 & 2.1       & $-$3.17         & +0.64          &  $-$0.85   & 2    & 21, 22                     &  0.49 & -- & 0.63   & 0.95\\  
\hline
\hline
\end{tabular}
\caption{Overview of literature data and derived probabilities for
  binarity for the targeted sample of CEMP-no stars in this
  work. Shown here are the literature values for the V-magnitude of
  the stars, derived Teff and log(g) (two possible solutions are given
  for 53327-2044-515), [Fe/H], [C/Fe] and [Ba/Fe]. The subsequent
  number of radial velocity literature measurements deviates slightly from the similar compilation of \citet{Norris13b} for a few stars. These differences
  arise because we count all individual measurements (also if they are on the same or adjecent days) as long as the velocities per
  observation are given separately in the relevant literature. The
  last four columns show the derived probabilities that the observed
  scatter in velocities is due to measurement errors (see text for
  details) for the data in the literature and this work both
  separately and combined. The last column shows the probability for
  the combined dataset, but inflating $\sigma_{vr_{i}}$ by a factor of
  3. References: 1 = \citet{Yong13a}; 2 = \citet{Ito09}, 3 =
  \citet{Norris10}, 4 = \citet{Cohen08}, 5 = \citet{Cohen06}, 6 =
  \citet{Frebel07a}, 7 = \citet{Honda04}, 8 = \citet{Lai08}, 9 =
  \citet{Cayrel04}, 10 = \citet{Spite05}, 11 = \citet{Francois07}, 12
  = \citet{Norris97b}, 13 = \citet{Norris13a}, 14 = \citet{Carney03},
  15 = \citet{Barklem05}, 16 = \citet{Mcwilliam95a}, 17 =
  \citet{Mcwilliam95b}, 18 = \citet{Norris01}, 19 =
  \citet{Depagne02},20 = \citet{Preston01}, 21 = \citet{Aoki13}, 22 =
  SDSS SSPP DR9, 23 = \citet{Cohen13} \label{tab:overview}}
}
\end{table*}
 
\section{Data}\label{sec:data}

Radial velocity measurements were obtained using the High Resolution
Spectrograph \citep[HRS,][]{Tull98} with resolving power R=18000 on the Hobby-Eberly
Telescope \citep[HET,][]{Ramsey98} from January to August 2013, as part of normal queue
observing \citep{Shetrone07}, after which the HET was taken
offline for installation of new instrumentation. During this period,
all extremely metal-poor CEMP-no stars as compiled by
\citet{Norris13b} in reach of HET were targeted at least twice. This
sample is restricted to CEMP-no stars that additionally have
[Fe/H]$<$--3.0, with one exception (CS~22878-027). We note that
53327-2044-515 will have an upper limit of [Ba/Fe] $<$ +0.34 if the star is on
the main-sequence and thus might or might not qualify the restrictions
for CEMP-no stars. In addition to this sample, we
targeted four new extremely metal-poor targets with CEMP-no-like
abundance patterns, as published in
\citet{Aoki13}. Although here we also note that two of these targets stars are on the border
of the CEMP-no star definition, because of their [Ba/Fe] determination
that is slightly above the solar ratio (although clearly not enhanced
to the [Ba/Fe]=+1 level of CEMP-s stars). J2206-0925 has
[C/Fe] = +0.64, which is slightly lower than the often used limit of
+0.70 for carbon-rich stars. An
overview of the literature stellar parameters and abundance ratios for these targets is presented in Table \ref{tab:overview}. 

For our targets, we calculated exposure times to obtain a minimum
S/N of $\sim$20. Directly after each target a Th-Ar exposure was taken and on nearly every night (weather permitting) a radial velocity standard was observed during twilight. Radial velocities were computed by
means of cross-correlating a synthesized extremely metal-poor
carbon-enhanced spectrum with the observed spectrum and
standard, using the velocity of the standard as
calibration for the zero point. The choice of a zero-point based on radial velocity standards is strengthened by a clear observed trend of the radial velocity standards' offsets with their literature values versus the observation date. The total magnitude of this variation over the full observing period is 2.7 \kms \  with small scatter. We are uncertain about its origin. However, since our radial velocity standards are observed at a similar time and comparable air-mass to our targets, any systematic effects that could be the cause of this offset in telescope, instrument or analysis are not propagated in our results.

To obtain the 1-$\sigma$ error bars on the measurements presented in this paper
we added in quadrature the individual 1-$\sigma$ errors in the radial
velocity cross-correlations with each spectrum and a subsequent error
floor of 0.26 \kms. This error floor is based on the rms variations
in all standard stars once corrected for the aforementioned variation with the date
of the observation by a simple 4th order polynomial fit. The final heliocentric velocities
and their errors for our observations are presented in Table
\ref{tab:results}. 

\begin{table}
\begin{tabular}{|l|r|r|r|}
\hline
{Star} &  {HJD}   &  {v$_{r}$} &  {$\sigma_{vr_{i}}$} \\
 &  {$-2456000$}  &  ($\kms$) &  ($\kms$) \\
\hline
\hline
53327-2044-515 & 487.93494 & $-$199.12 &  1.39 \\
53327-2044-515 & 510.87326 & $-$205.21 &  2.60 \\
53327-2044-515 & 522.84846 & $-$198.70 &  4.39 \\
\hline
BD~+44-493  & 330.61789 & $-$150.26 &  0.37 \\
BD~+44-493  & 339.62141 & $-$150.08 &  0.37 \\
BD~+44-493  & 486.95551 & $-$150.17 &  0.30 \\
BD~+44-493  & 508.91578 & $-$149.94 &  0.32 \\
\hline
BS~16929-005 &  341.81244 &  $-$50.62 & 0.26 \\
BS~16929-005 &  366.74166 & $-$50.61 & 0.77 \\
BS~16929-005 &  396.88416 & $-$50.48 & 0.29 \\
BS~16929-005 &  469.68353 & $-$50.44 & 0.95 \\
BS~16929-005 &  479.65853 & $-$50.18 & 0.26 \\
\hline
CS~22878-027 &  384.87258 & $-$91.41 & 0.36 \\
CS~22878-027 &  389.85421 & $-$91.99 & 0.53 \\
CS~22878-027 &  420.76769 & $-$91.14 & 0.57 \\
CS~22878-027 &  455.68007 & $-$91.16 & 0.26 \\
CS~22878-027 &  480.77878 & $-$91.64 & 0.66 \\
CS~22878-027 &  507.70591 & $-$91.76 & 0.29 \\
\hline
CS~22949-037 & 485.92583 & $-$126.00 & 0.36 \\
CS~22949-037 & 506.86908 & $-$125.84 & 0.27 \\
\hline
CS~22957-027 & 487.95236 &  $-$60.79 & 0.53 \\
CS~22957-027 & 505.94683 &  $-$60.75 & 0.41 \\
\hline
CS~29502-092 & 458.95257 & $-$66.93 & 0.40  \\
CS~29502-092 & 478.90277 & $-$66.72 & 0.35  \\
CS~29502-092 & 506.89918 & $-$66.73 & 0.54  \\
\hline
HE~1150-0428 & 339.88474 & 47.93 &  0.28  \\
HE~1150-0428 & 354.83867 & 43.85 &  0.29  \\
HE~1150-0428 & 367.80309 & 41.09 &  0.52  \\
HE~1150-0428 & 387.76986 & 36.88 &  0.26  \\
HE~1150-0428 & 413.69469 & 35.79 &  0.33  \\
\hline
HE~1300+0157 & 330.89601 & 74.86 &  0.37  \\
HE~1300+0157 & 355.83157 & 74.29 &  0.55  \\
HE~1300+0157 & 368.90131 & 74.42 &  0.45  \\
HE~1300+0157 & 374.77741 & 73.60 &  0.84  \\
HE~1300+0157 & 413.78680 & 75.19 &  0.48  \\
HE~1300+0157 & 459.66750 & 74.39 &  0.29  \\
\hline
HE~1506-0113 & 355.93837 & $-$80.12 & 0.27 \\
HE~1506-0113 & 369.89926 & $-$79.55 & 0.29 \\
HE~1506-0113 & 384.94103 & $-$80.25 & 0.59 \\
HE~1506-0113 & 416.85083 & $-$82.07 & 0.86 \\
HE~1506-0113 & 455.74001 & $-$86.39 & 0.28 \\
HE~1506-0113 & 474.67805 & $-$88.93 & 0.29 \\
HE~1506-0113 & 487.65958 & $-$90.74 & 0.33 \\
\hline
SDSS~J1422+0031  & 442.65168 & $-$124.63 & 0.50 \\
\hline
SDSS~J1613+5309  & 442.75102 & 0.07    & 0.26 \\
SDSS~J1613+5309  & 512.70831 & $-$0.60    & 0.88 \\
\hline
SDSS~J1746+2455  & 442.74986 & 78.03  & 0.28 \\
SDSS~J1746+2455  & 455.70036 & 79.52  & 1.80 \\
SDSS~J1746+2455  & 487.82824 & 78.56  & 0.70 \\
\hline
SDSS~J2206-0925 & 481.91425 & 14.66 &  0.26 \\
\hline
Segue 1-7 & 339.69733 & 204.01 & 1.70 \\
Segue 1-7 & 367.61593 & 205.26 & 3.06 \\
Segue 1-7 & 391.74756 & 205.04 & 0.27 \\
\hline
\hline
\end{tabular}
\caption{Data added in this program \label{tab:results}}
\end{table}

\section{Results}\label{sec:radvel}

\subsection{Radial velocity variations}

\begin{figure*}
\centering
\includegraphics[width=0.8\linewidth]{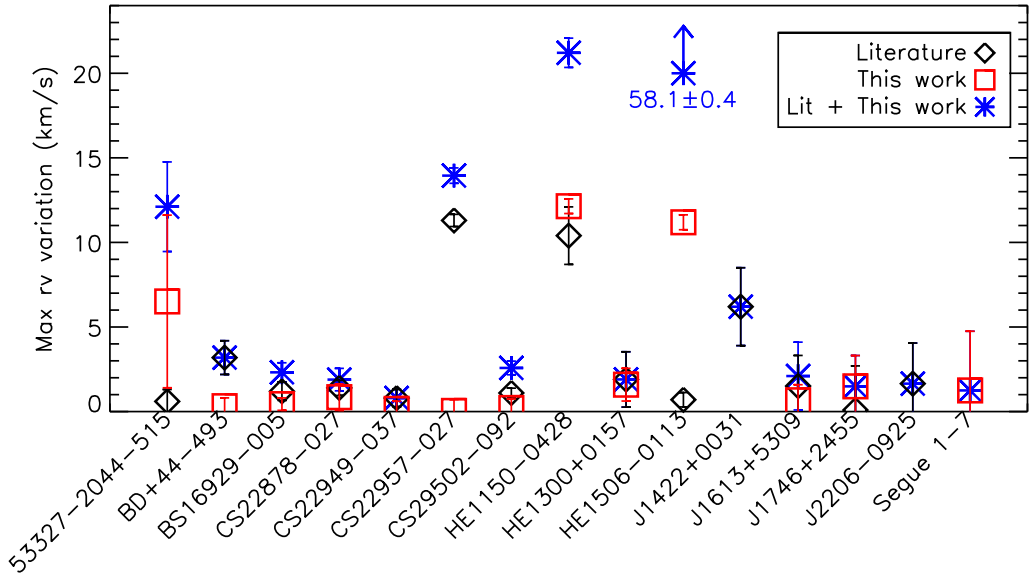}
\caption{Maximum velocity variation for each of the stars observed in this work. The largest difference between two measurements is shown for the data presented in Table \ref{tab:results} (red squares), literature data only (black diamonds) and both datasets combined (blue asterisks). If one of the subsets had only one measurement, the plotting symbol for that subset is omitted. Error bars reflect the measurement errors on both measurements involved. The velocity difference between our data and the literature measurement for HE~1506-0113 is beyond the scale of this figure, its value is shown in the panel instead. \label{fig:K_ll}}
\end{figure*}

In Figure \ref{fig:K_ll} we show the maximum velocity variation between two measurements for all stars observed in this program, taking into account the data obtained in this work and literature data both separately and combined. Two stars very clearly stand out: both HE~1150-0428 and HE~1506-0113 are varying by more than 10 \kms \  over the 74 and 132 days these stars were observed with the HET. Moreover, the currently observed heliocentric velocity of HE~1506-0113 differs by over 57 \kms \  from the measured velocity by \citet{Norris13a}.

In addition to binarity, flows, pulsations and inhomogeneities on the stellar
surface can also present themselves in observed radial velocity
variations. Considering the precision of this study, this so-called
radial velocity ``jitter'' -- although theoretically poorly understood -- is
expected to (only) manifest
itself clearly at the tip of the red giant branch
\citep[e.g.][]{Gunn79, Setiawan03, Wright05,Carney08}. \citet{Carney08} find
in their monitoring program of metal-poor stars that jitter
is present in most stars with $M_{V} < -2.0$ and in a significant
number of stars with $M_{V} < -1.4$. Based on their results, they cannot exclude that velocity jitter is not contributing at lower
magnitudes as well, but they conclude that such an assumption is reasonable. From the spectroscopically
derived parameters presented in Table \ref{tab:overview}, we do not expect any of
our stars to be in this regime, and we will therefore treat radial velocity
variations as attributed to binarity for the remainder of this
paper.

For several targets, additional data is needed because the current variations are small, or rely heavily on the accuracy and comparison between literature data taken at different telescopes and analyzed by different teams. This is true for example for J1422+0031, for which the current binarity is based on three measurements and 53327-2044-515, that only shows clear evidence for binarity when our data set and the literature are combined. We continue to pursue follow-up observations, and anticipate to be able to constrain these particular cases better in future work.   

\subsection{Quantifying binarity}

Following \citet{Lucatello05}, we first quantify our results by calculating the $\chi^{2}$ value for the radial velocity distribution.      
\begin{equation}\label{chisq}
\chi^{2} = \sum_{i=1}^{n}\bigl(\frac{v_{{r}_{i}}-\bar{v_{r}}}{\sigma_{v_{{r}_{i}}}}\bigr)^{2} 
\end{equation}
We subsequently evaluate the probability that the radial velocities observed are compatible within the measurement errors expressed as $p(\chi^{2}|f)$, where $f$ is the number of degrees of freedom. A small $p$-value thus indicates a low probability that the observed scatter in velocities is due to measurement errors, and points to an additional source of velocity variability. However, instead of a simple mean for all observations, we are using a mean value that is weighted by the observational errors. We find that the use of a weighted mean is critical for obtaining the correct $\chi^{2}$ in datasets where the magnitude of the errors varies substantially from observation to observation for the same star (for instance, because different datasets are combined or exposure times vary). Therefore we use:
\begin{equation}\label{weight}
\bar{v_{r}} = \frac{\sum_{i=1}^{n} w_{i}v_{{r}_{i}}}{\sum_{i=1}^{n} w_{i}}\textnormal{ and}\ w_{i}=\frac{1}{\sigma_{v_{{r}_{i}}}^{2}} 
\end{equation}
$p$-Values for our dataset, as well as for the literature and combined datasets, are given in Table \ref{tab:overview}. We note that \citet{Lucatello05} include a multiplication by a factor 3 in all $\sigma$-values. This extra factor is motivated by the work of \citet{Preston01} on multiple observations of giant stars, but is nevertheless quite arbitrary. In this work, we use the 1$\sigma$ errors, except in the last column of the table, where we have included this extra factor of 3. From a comparison between the results with 1$\sigma$ and 3$\sigma$ errors, it is clear that the treatment of errors can be a driving parameter for the derivation of the binary fraction of the population.

\subsection{Period analysis}

\begin{figure*}
\centering
\includegraphics[width=0.5\linewidth]{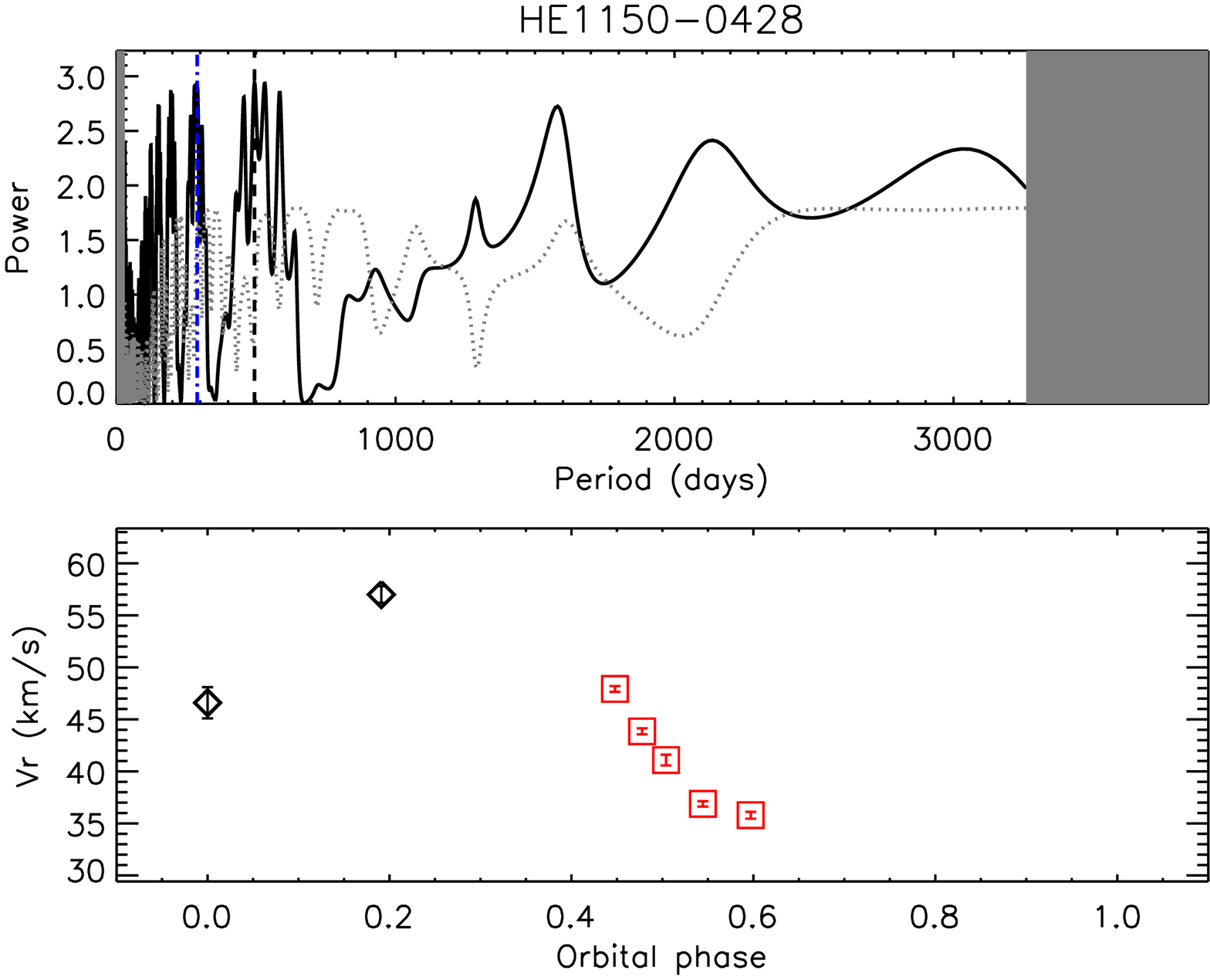}\includegraphics[width=0.5\linewidth]{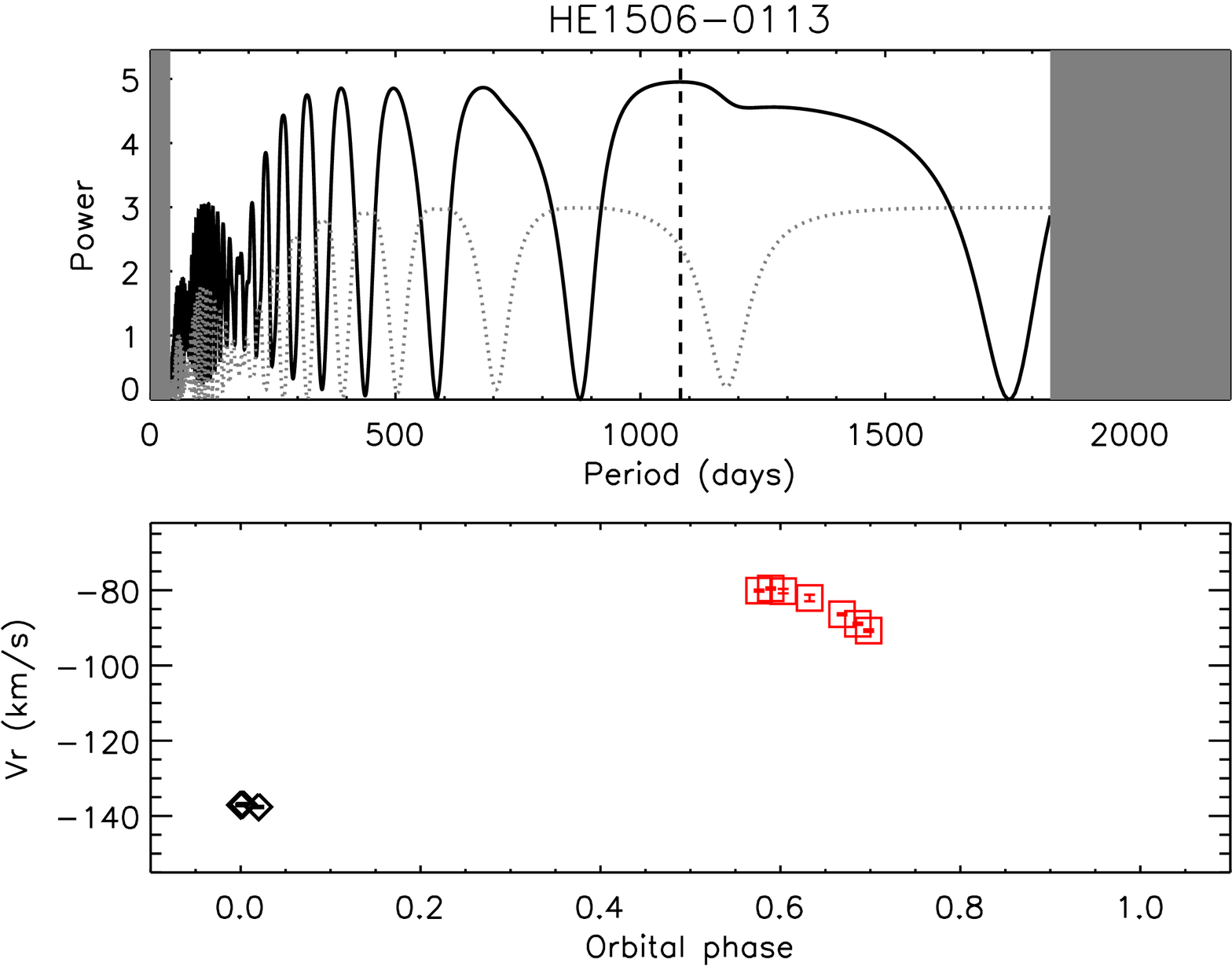}
\caption{Periodograms of HE~1150-0428 and HE~1506-0113 based on all
data, both literature and from this team (black diamonds and red squares in the bottom panel). The black line in the top panel indicates
the power of the period solution. The same analysis is subsequently
performed on a constant function with exact same
sampling as the real data to check for any peaks that might be
spurious results due to the sampling alone (dotted grey line). Various
different periods are still very likely with the current data ranging
from several weeks or months to years. The most likely orbital period
from this analysis is highlighted by a dashed vertical black line. In
the top left panel the blue dot-dashed line indicates a period of 289 days
as found by Andersen et al., in preparation for the
same star. Orbital phases for the velocity data for our most likely possibile orbit period are shown in the bottom panel. \label{fig:perio}}
\end{figure*}

A second step, after detecting variability, is to constrain the
parameters of the stars' binary orbits, and thus gain a deeper physical insight in
the possible pollution of these stars by companions. Shown in Figure
\ref{fig:perio} are periodograms for HE~1150-0428 and HE~1506-0113 from
a period analysis of the (unevenly sampled) time series of radial
velocity monitoring using Equation 1. from \citet{Horne86}. This
method is equivalent to a least square-fitting of sine curves to the
data. The minimum period sampled is taken as the average sampling frequency for the five radial velocity datapoints closest in time (the minimum amount required to do a period analysis),
the time between the first and last data
point is the maximum period sampled. The peaks are checked against false positives by additionally
performing the same analysis on a constant function with exact same
sampling as the real data (see dotted lines in Figure \ref{fig:perio}); any
peaks that correspond between the filled and dotted lines might be
spurious results due to the sampling alone. 

Various different periods, ranging from several weeks or months to years, are still very likely with the current data for both stars. More data with a good cadence will be required to more accurately define their periods. We note that \citet{Cohen13} mention that a period of 289 days for HE~1150-0428 was found by Andersen et al., in preparation. Indeed, such a period would be consistent with our results, as highlighted by the blue vertical dot-dashed line in the bottom left panel of Figure \ref{fig:perio}.

\section{Constraining binary fraction and binary periods for carbon-enhanced metal-poor stars}\label{sec:sims}

We use tailored Monte Carlo simulations combined with a maximum likelihood analysis to constrain the binary fraction and binary periods for various types of carbon-enhanced stars, and investigate their nature. The binary fraction and period are investigated simultaneously because they are degenerate. If only a relatively small number of stars shows radial velocity variations, this can point towards both a lower binary fraction or a higher average period; in the latter case the observational cadence -- which is often on the order of years -- might not be able to pick up any signal of variations with enough accuracy. To test both parameters, we draw 10,000 realizations of radial velocity datasets from a simulation with a certain combination of binary fraction (0 to 100\%) and a cut-off period for these binaries. The mock datasets have the same cadence and are convolved with the same velocity errors as the observed datasets. 

Parameters for the orbits of the binary stars are randomly assigned
using constraints derived from the study of \citet{Duquennoy91} in the
Local Neighbourhood. We stress here that it is unclear if indeed our
data will follow the same characteristics as the Solar Neighbourhood
binaries, but -- lacking any other constraints -- it seems a
reasonable assumption. We select stellar mass ratios for the binary
pair from \citet{Duquennoy91}, but constrain the mass of the star
observed $M_{1}= 0.8\msun$, consistent with an old age (sub)giant. A
comparison of the stellar parameters in Table \ref{tab:overview}
confirms that this is indeed the expected mass for a large majority of
the stars if they are of old age ($\sim13$ Gyr). Eccentricities are selected from a thermal distribution. The inclination $i$ and the longitude at the ascending node $\omega$ are picked from a uniform distribution. The initial phase $\nu_{0}$ is randomly selected from a distribution in accordance with the selected eccentricity of the orbit (i.e., uniformly for a circular orbit, but more likely to be at apocentre for an highly elliptical orbit). 

The orbital periods for the binaries (if the binary fraction is not zero), $P$ in days, are characterized following \citet{Duquennoy91} by $\overline{\textnormal{log}{P}}=4.8$ and $\sigma_{logP}=2.3$. We introduce an upper and lower limit to the periods the mock stars are allowed to have. The applied lower limit excludes really short period binaries from our simulations; we set it at the 2$\sigma$ level of $\sim$1 day and this is kept the same for all simulations. The upper limit is consequently varied as a second free parameter (besides the binary fraction), and ranges from almost a year up to the $2\sigma$ upper level for the Solar Neighbourhood distribution of $\sim7$ million years, in total covering six orders of magnitude.

We use a maximum likelihood analysis to return the most likely parameters characterizing the observed distribution. We compare the modeled distributions, $M$, to our datasets, $D$, by constructing histograms of the mutual independent velocity variations for each star, $i$ and each observation $t$, using the first measured velocity as a reference point.

\begin{equation}
\Delta v_{i,t} =  \frac{|v_{r_{i,0}} - v_{r_{i,t}}|}{(\sigma^{2}_{v_{r_{i,0}}} + \sigma^{2}_{v_{r_{i,t}}})^{\frac{1}{2}}}
\end{equation}

\begin{figure}
\includegraphics[width=0.9\linewidth]{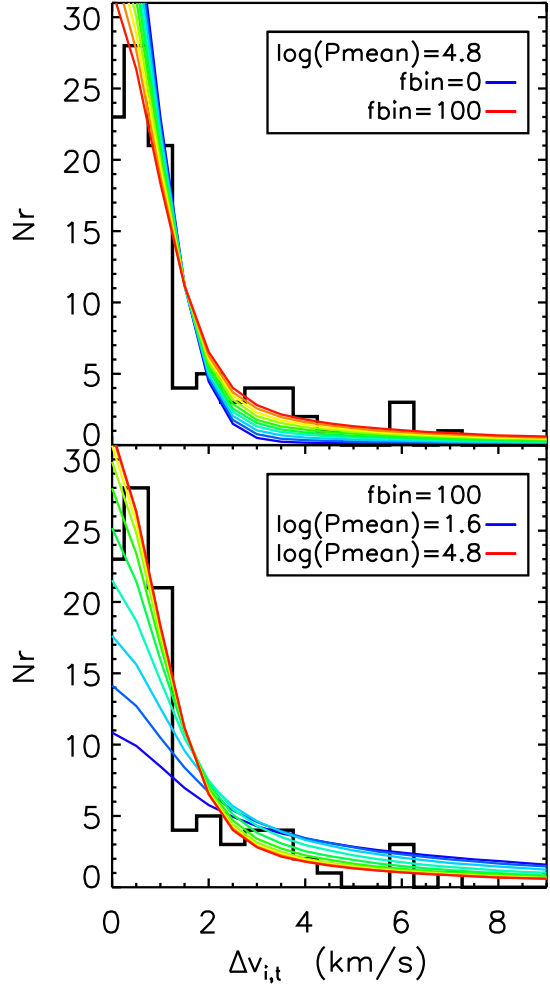}
\caption{$\Delta v_{i,t}$ for the full CEMP-no dataset (literature +
  this work) overplotted with the normalized curve for 10,000
  simulated realizations with similar cadence and errors. In the
  simulations the binary fraction and maximum period are left
  variable. The top panel shows how -- at a Solar Neighbourhood value
  for P$_{\textnormal{max}}$ (and therefore
  P$_{\textnormal{mean}}$ = 4.8) -- the distribution changes from no binaries
  (blue) to 100\% (red). The bottom panel shows how -- at a fixed
  binary fraction of 100\% -- the distribution changes from a value for
  P$_{\textnormal{max}}$ 2$\sigma$ away from the solar mean (10$^{9.4}$ days, red) to our adopted smallest maximum value of 312 days (blue).   \label{fig:param}}
\end{figure}

Figure \ref{fig:param} shows how the two parameters that we allow
to vary -- the binary fraction, $f_{bin}$, and maximum period, $P_{max}$ -- change the
distribution of 10,000 mock datasets in this space. Varying any of the
two parameters will change the peak height and the tail of the
distribution. It is clear that some degeneracy between the two
parameters is met. Subsequently, we determine the most likely values for our variables by applying Bayes' theorem and assuming a uniform prior probability of the parameters. Assuming Poisson uncertainties in each separate histogram bin, we can express the relative posterior probability distribution for each combination of $f_{bin}$ and $P_{max}$ considered as

\begin{equation}
p(f_{bin}, P_{max}|D,M) \propto \prod^{N}_{b=1} \frac{m^{d_{b}}_{b} \textnormal{e}^{-m_{b}}}{d_{b}!},
\end{equation}

where $d_{b}$ is the number of velocities measured with a certain
variation in the dataset, and $m_{b}$ the expected number of
observations in that same velocity variation bin from the model,
normalized to the same number of total observations as in the
dataset. For practical reasons we approximate the Poisson distribution by a Gaussian distribution with mean $m_{b}$ and variance $m_{b}$ for values of $m_{b}$ greater than 16. Each individual probability, $p(f_{bin}, P_{max}|D,M)$, is
subsequently normalized by the sum of all probabilities across our
model parameters. 

As a simple test, we apply our method on the dataset of metal-poor stars by \citet{Carney03}. We find that an initial modeled binary fraction of 60\% with a maximum period of $\textnormal{log}{P}=9.4$ \citep[corresponding to the 2$\sigma$ upper level, thus ensuring that the average period is identical to that derived by][in the Local Neighbourhood]{Duquennoy91} convolved with the cadence and errors of their dataset returns 34\% detectable binaries with $p(\chi^{2}|f)<0.01$, in good accord with their findings in the data (32\%).

\subsection{A maximum likelihood analysis for the CEMP-no sample}
To obtain a complete set of CEMP-no velocity monitoring, we combine all data presented in Table \ref{tab:overview} with the stars published in \citet{Norris13b}, out of reach of HET, but for which multiple radial velocities are published \citep{Aoki04,Aoki06,Norris07}. For all literature data we exclude measurements that are published without an error, or do not mention the date of the observation. 
\begin{figure}
\includegraphics[width=0.84\linewidth]{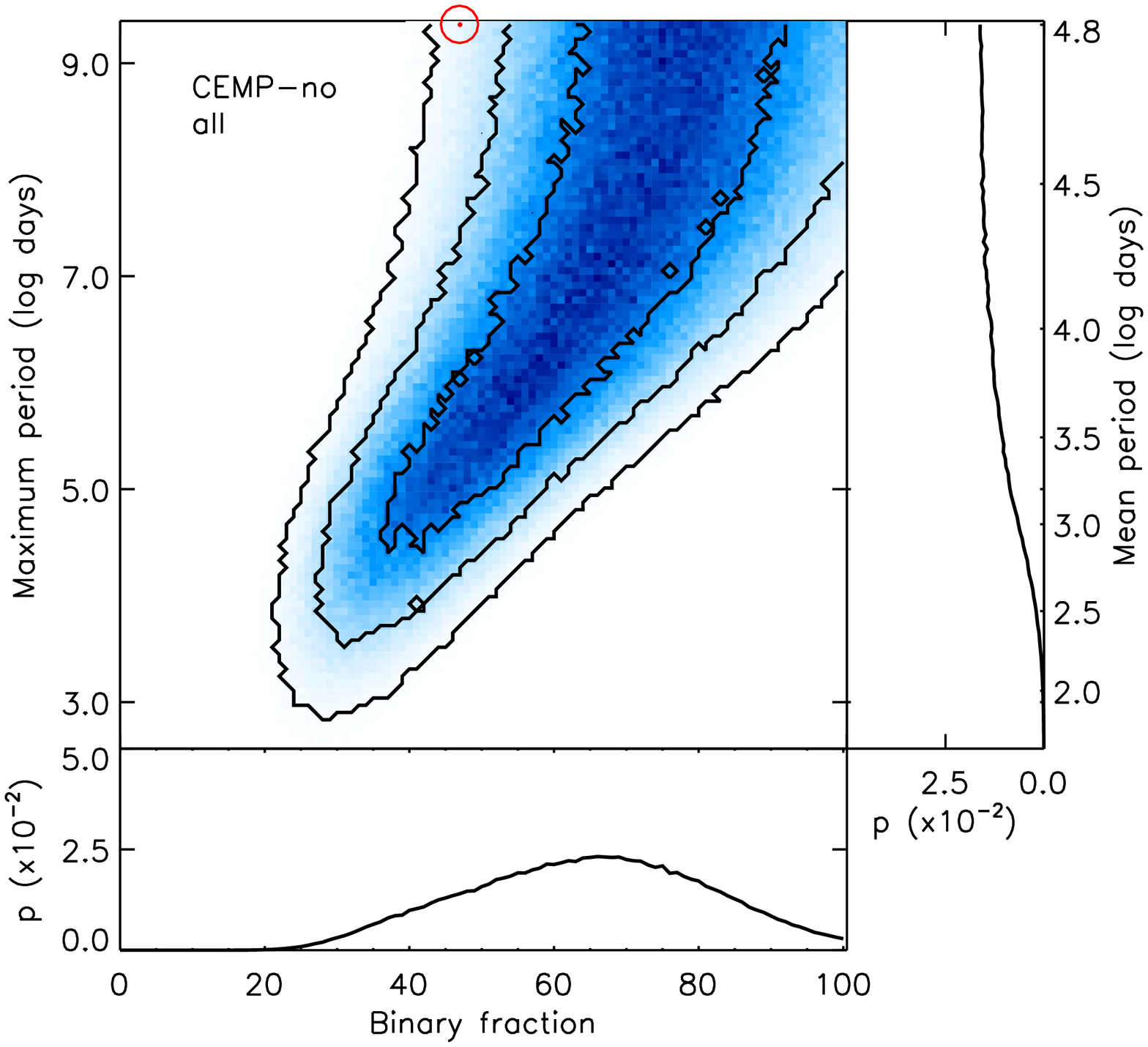}\includegraphics[width=0.15\linewidth]{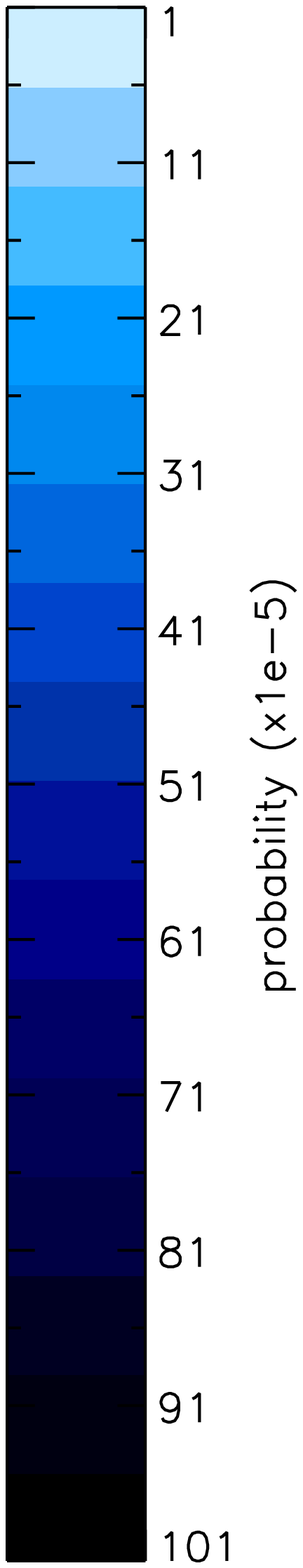}
\caption{The relative posterior probability (see text for details) of each of the combinations of
  binary fraction and maximum period for the full sample of CEMP-no
  stars. Contour levels are drawn at the 1-, 2- and
  3-$\sigma$ levels. The probabilities marginalized over one of the free
  parameters are shown in an extra panel next to and below the contour
  level plots. The binarity fraction and period distribution in the Solar Neighbourhood among solar type stars is marked with a red solar symbol. \label{fig:simcempno}}
\end{figure}

Figure \ref{fig:simcempno} shows the results of the posterior
probability distribution as a function of the free parameters
$f_{bin}$ and $P_{max}$. As seen from this figure, a high amount of degeneracy is indeed met in constraining the binary fraction and the binary periods.  The present data for the CEMP-no stars are only very marginally compatible with the Solar Neighbourhood binary properties among solar type stars \citep[marked with a solar symbol in Figure \ref{fig:simcempno}, properties are taken from \citet{Duquennoy91} and][]{Raghavan10}.

\begin{figure*}
\includegraphics[width=0.45\linewidth]{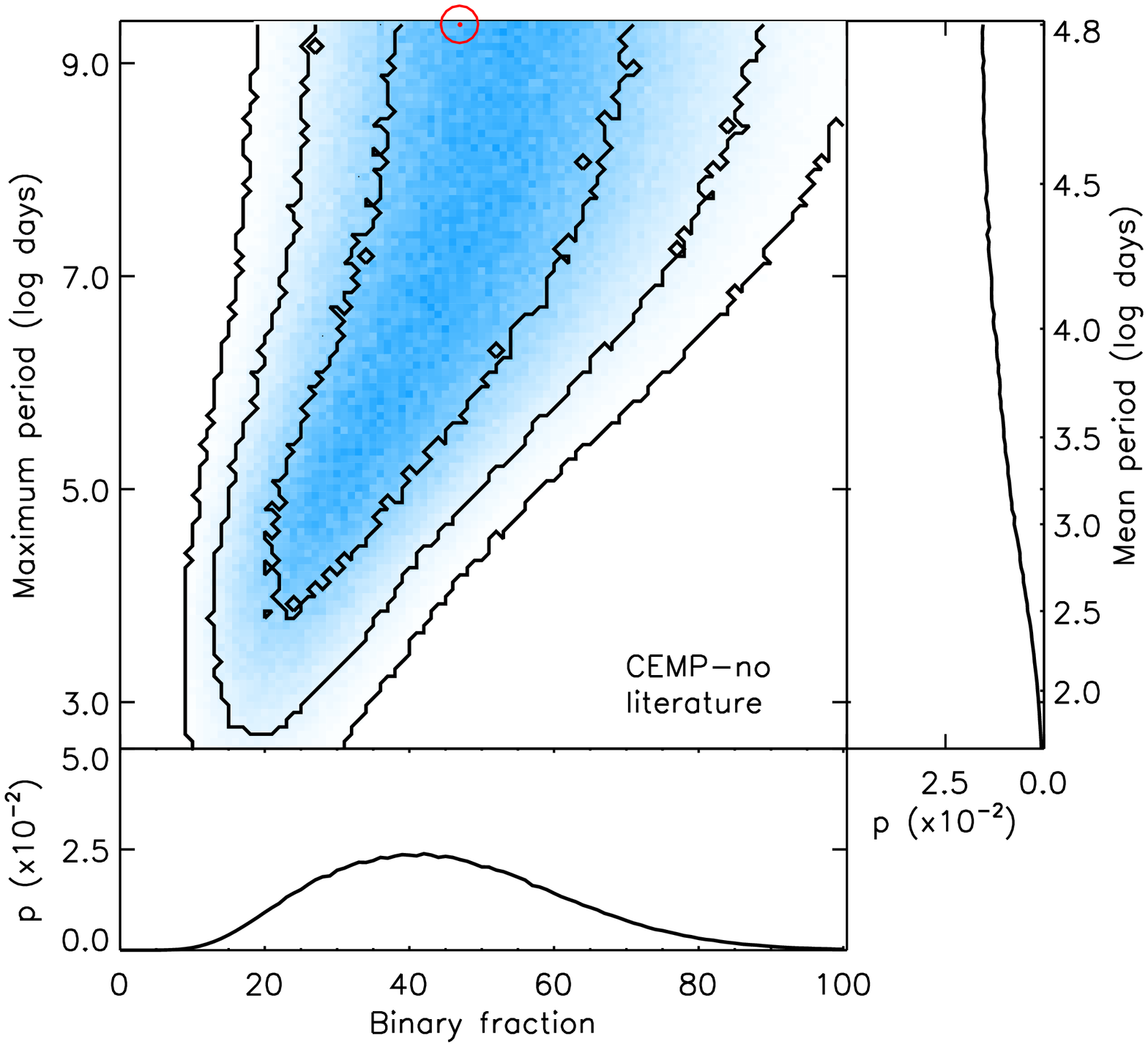}\includegraphics[width=0.45\linewidth]{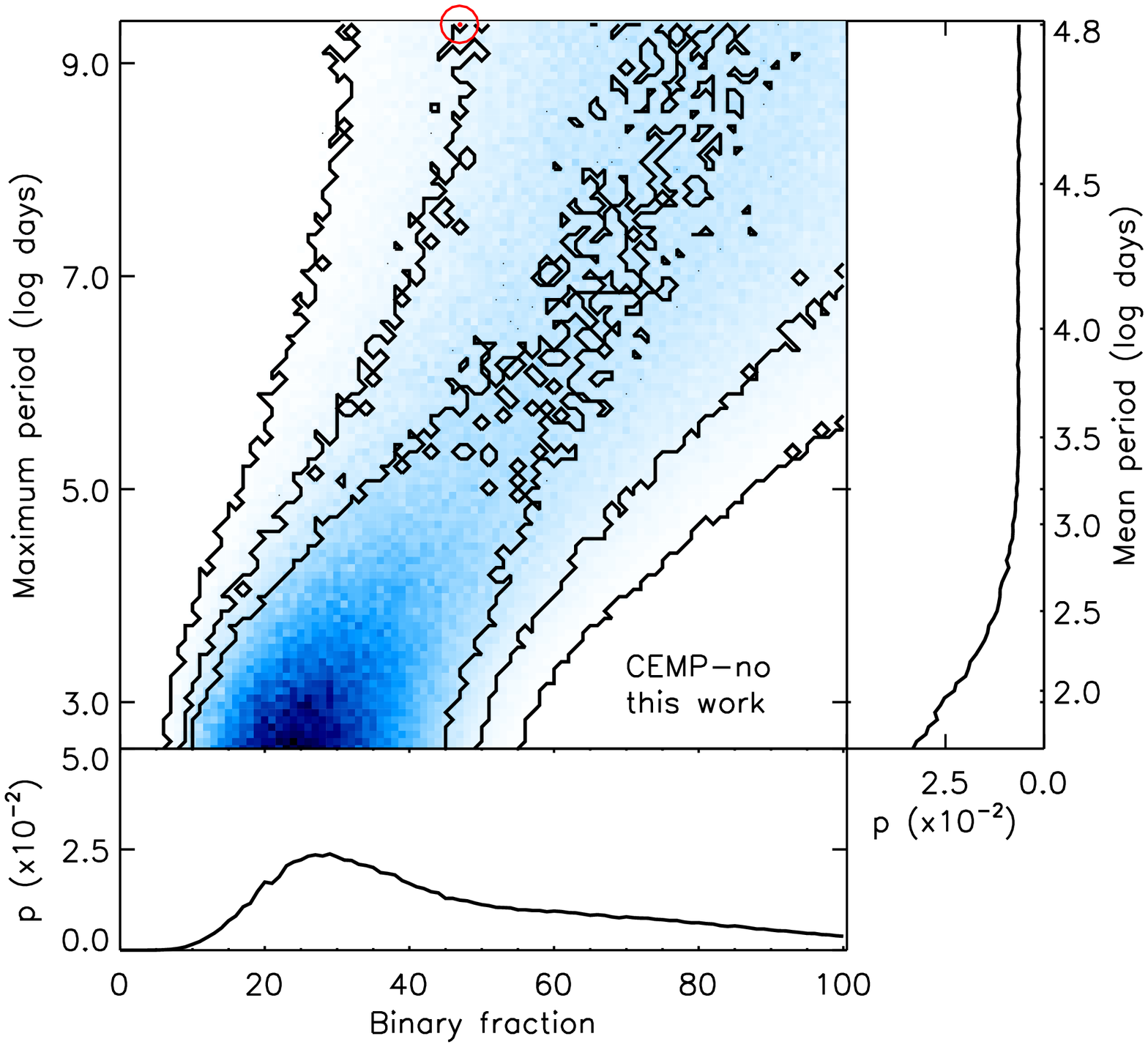}\includegraphics[width=0.082\linewidth]{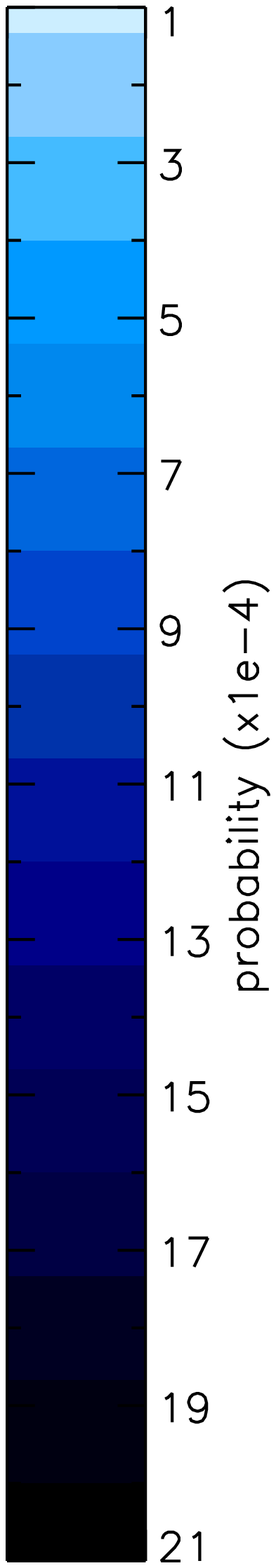}\\
\caption{The relative posterior probability (see text for details) of each of the combinations of
  binary fraction and maximum period for the CEMP-no sample, divided into
  the data from this work and the data from literature. As in Figure \ref{fig:simcempno},
 contour levels are drawn at the 1-, 2- and
  3-$\sigma$ levels, and the probabilities added over one of the free
  parameters are shown in an extra panel next to and below the contour
  level plots. The binarity fraction and period distribution in the Solar Neighbourhood among solar type stars is marked with a red solar symbol.\label{fig:simsustheir}}
\end{figure*}

In Figure \ref{fig:simsustheir} we split the sample of CEMP-no stars
in the data from this work and the literature data. Although both
subsamples show a similar trend in their 2-$\sigma$ contours, it is
interesting to see that the best solutions populate a different part
of the diagram in both subsamples. Note that the probabilities
indicate the best combination of both free parameters $f_{bin}$ and
$P_{max}$ for the full dataset, and not a likelihood of the properties
of any individual star. It is therefore not to be expected that a
combination of the two panels of Figure \ref{fig:simsustheir} would
result in Figure \ref{fig:simcempno}. 

In the subset of data from this
work only, it seems most likely that a few stars are in close period
binaries, and the rest of the sample consists of single stars. In the
literature dataset, the preferred solution points more toward a
solution in which many stars are in binary systems, but many of these
binary systems are wide binaries. As detailed above, these two
solutions can be degenerate. One important consideration is that our
cadence is too short to robustly detect long period
binaries and therefore will likely classify long-period binaries as
single stars. Another consideration that could drive the
offset is the treatment of errors. If in the literature dataset
systematic errors between the measurements are underestimated, for example, this
will result in many (single) stars showing small variations that are not recognized as being caused by errors and will push
the overall distribution towards more and wider binaries. A similar
mismatch would occur if the literature errors are correct, but our
errors are overestimated. With the current data in hand, it is difficult to
fully break the degeneracy. It is however clear from both subsamples
and the full dataset that there are clear binaries among these stars,
that a Solar Neighbourhood distribution is only marginally acceptable,
and that \textit{if} all these stars are in binaries, most of these
binaries will have to have very long periods.

\subsection{The CEMP-s and CH-star samples}
\begin{figure*}
\includegraphics[width=0.42\linewidth]{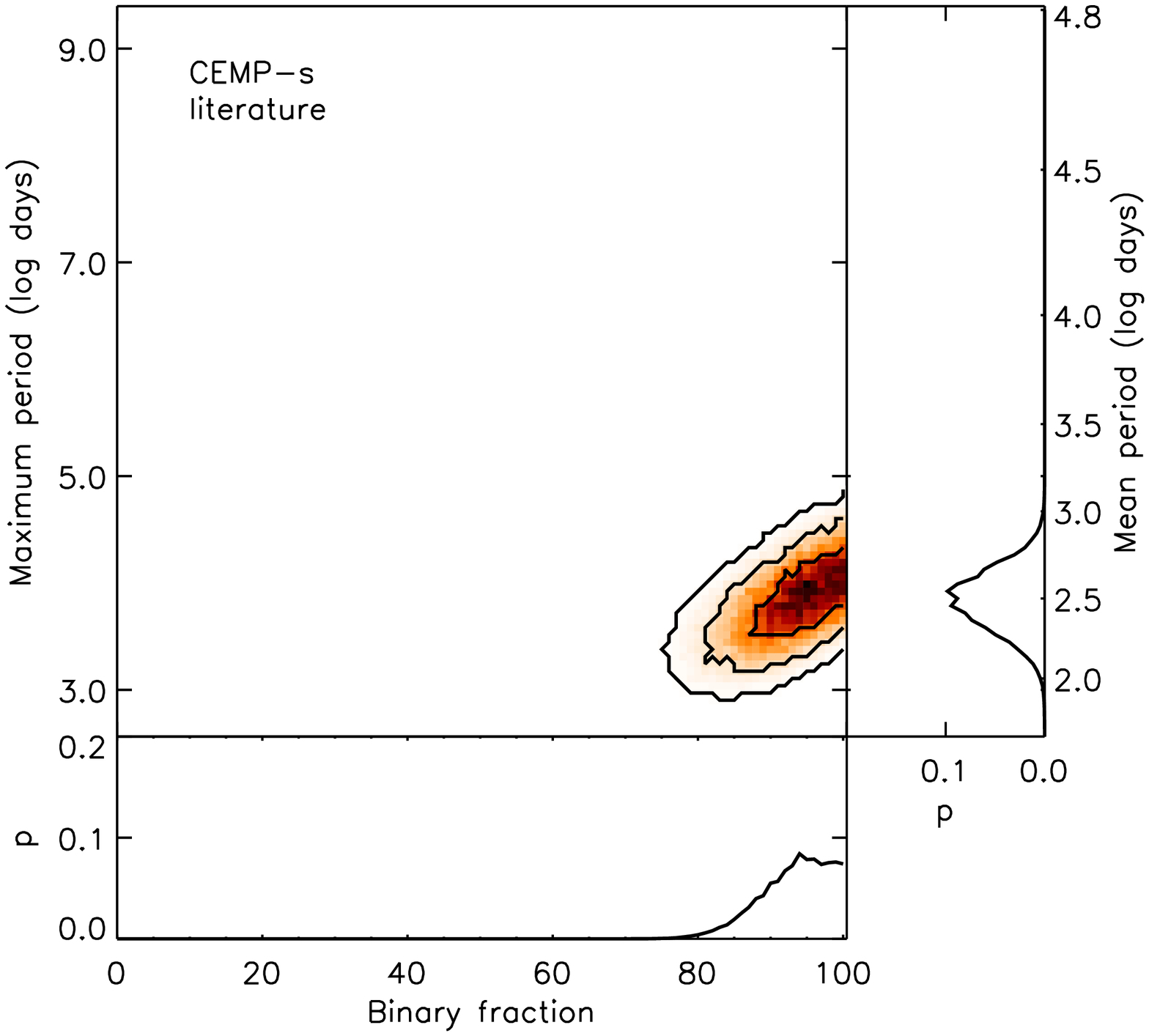}\includegraphics[width=0.075\linewidth]{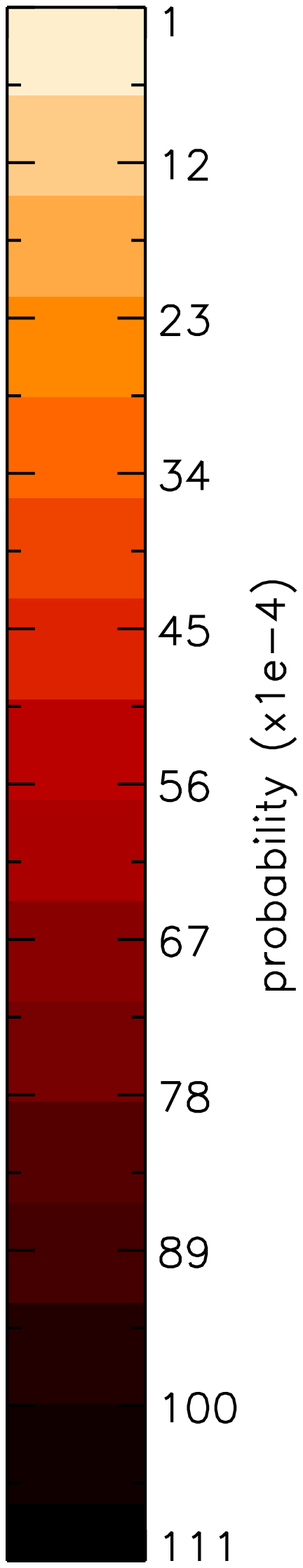}\includegraphics[width=0.42\linewidth]{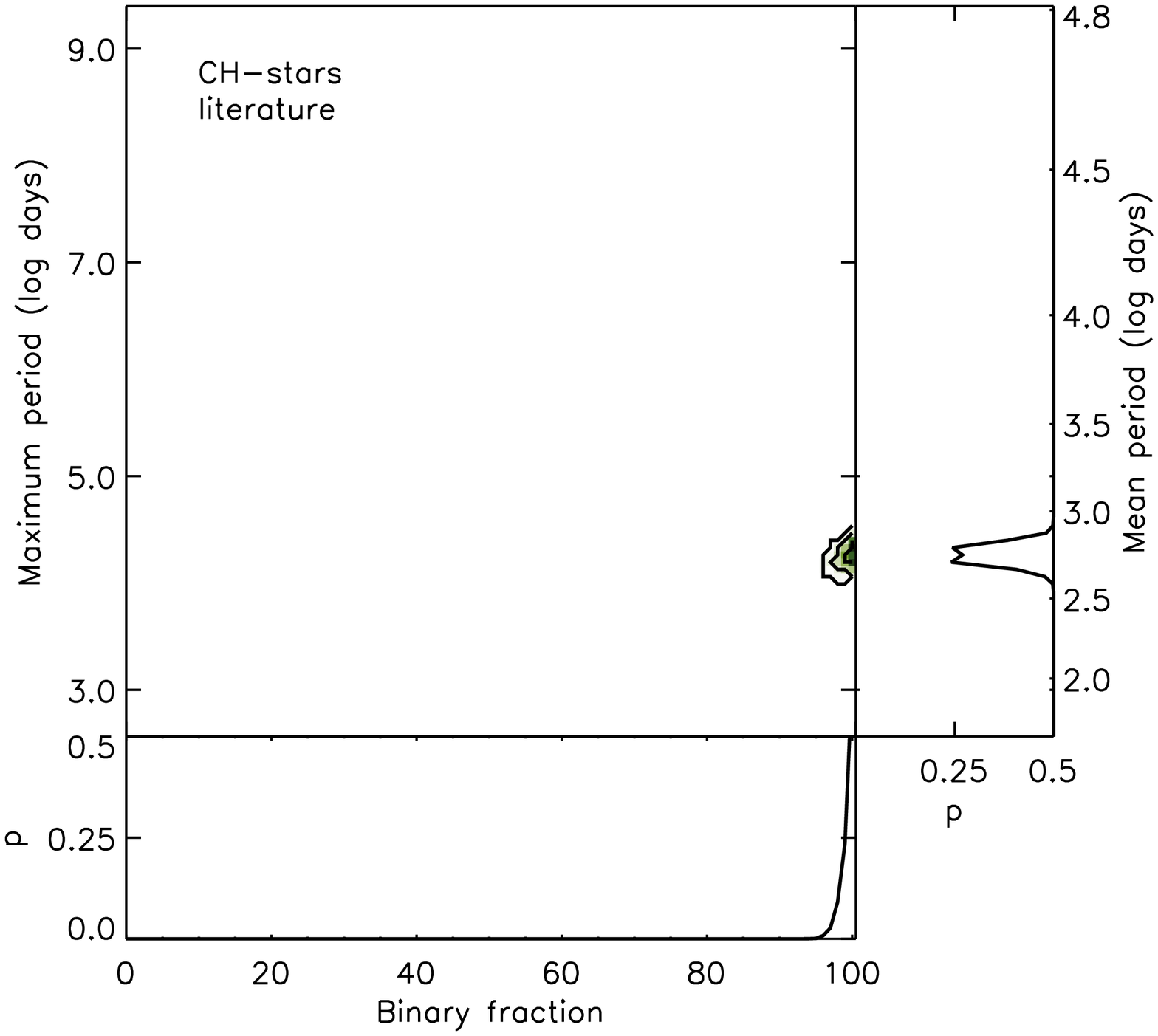}\includegraphics[width=0.075\linewidth]{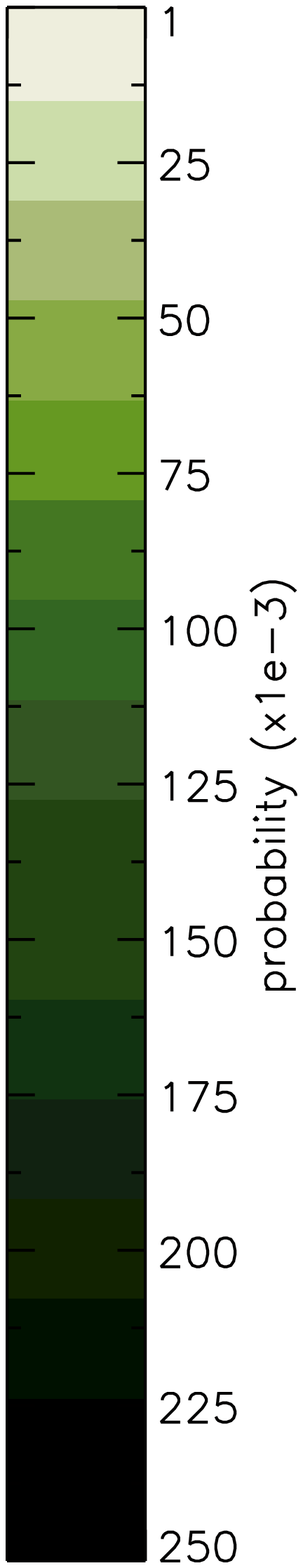}
\caption{The relative posterior probability (see text for details) of each of the combinations of
  binary fraction and maximum period for the full literature samples of CEMP-s stars \citep{Lucatello05}, and CH-stars \citep{McClure90,McClure97a}. Contour levels are drawn at the 1-, 2- and
  3-$\sigma$ levels. The probabilities added over one of the free
  parameters are shown in an extra panel next to and below the contour
  level plots. \label{fig:simcemps}}
\end{figure*}

In Figure \ref{fig:simcemps} we apply the same method to two other
related samples of stars with extended radial velocity monitoring. First of all we take the
CEMP-s sample from \citet{Lucatello05}, which consists of their data combined with radial velocity measurements from \citet{McClure90,Norris97a,Hill00,Preston00,Preston01,Aoki01,Aoki02a,Aoki02b,Sneden03,VanEck03,Lucatello03,Cohen03} and \citet{Barbuy05}. The number of observations and the length of the baselines for
the CEMP-s velocity monitoring is fairly comparable to
observational evidence of the CEMP-no stars, when the data in this work is added to the literature measurements. 

Secondly, we have also analyzed the radial velocity monitoring for the much higher metallicity CH-, sgCH- and Ba II-stars as gathered by
\citet{McClure90} and \citet{McClure97a}. These stars all show carbon
and s-process enhancements, but they were originally placed in
different classes based on their luminosity. For convenience, we will
in the remainder of this paper refer to this combined sample as
``CH-stars''. We note that three stars in their sample were classified
as CEMP-s stars by \citet{Lucatello05} and are present in both
samples. This fact illustrates the difficulty to draw sharp boundaries
between various classes that are only loosely defined. A better physical understanding of their origin will help to define these classes in a more robust way. 

Interestingly, the contours for both the CEMP-s
and CH-stars are very well defined, both in binary fraction and
in period. Our analysis shows with high probability that all, or
almost all, CH-stars and CEMP-s stars are in binaries, confirming the
conclusions of \citet{Lucatello05} for CEMP-s stars and \citet{McClure80, McClure83, McClure84, Jorissen88, McClure90, McClure97a, Jorissen98} for the CH-stars. In addition to these results in the literature, we also show convincingly that all these stars are in tighter
binaries than the average Solar Neighbourhood binary system. Our most
likely solution indicates in both cases a maximum period of
$\sim$10,000-20,000 days, corresponding to an average period of $\sim$400-600 days. 

Such relatively short periods are indeed expected in a scenario in
which mass transfer is the mechanism responsible for their abnormal
chemical pattern \citep[e.g.,][]{Han95}. However, despite detailed modelling efforts, it is poorly understood what the driving mechanism for mass transfer onto companion stars would be. It has been argued that pure Roche-lobe overflow from an AGB star is usually unstable, and will lead to negligible accretion \citep{Paczynski65,Ricker08}. Therefore, much effort has been focussed on accretion by stellar winds, or a scenario called ``wind Roche-lobe overflow (WRLOF)'' \citep{Mohamed07,deVal09,Abate13}. 
 As shown in \citet{Abate13}, various unknowns in the exact parameters
of the mass transfer will lead to different final orbits, due to
angular momentum loss in the process. Our results for these stars seem
in better qualitative agreement with their suite of models that
include significant angular momentum loss. These models are
characterized by a peak in the final periods somewhat higher than
$\sim$1000 days. Instead, their standard wind models are peaked at
periods significantly higher than our best-fit P$_{max}$
\citep{Izzard09}. However, we note that the shape of the period distribution
in their models shows a double peak, whereas we still assume a lognormal
distribution with a cutoff. We therefore caution against a strict comparison of our derived maximum
and mean to simulations and individual orbit solutions.

With this caveat in mind, our results
-- derived from the population as a whole -- are also in qualitative agreement
with the range of orbits found for several of these stars
individually. Within the total sample of nineteen stars, the ten
CEMP-s stars with orbital solutions have periods ranging from 3.4 to
4130 days, with an average period of 1200 days \citep[][and references
therein]{Lucatello05}. For the 32 stars monitored by
\citet{McClure90}, 24 stars have orbital solution and show a range of
periods from 80 to 4390 days and an average of 1650 days. Additionally, \citet{Jorissen98} present a comparison of orbital solutions from radial velocity monitoring results \citep[see also][]{Udry98a,Udry98b} of 93 binaries with (mild and strong) barium stars and stars classified as Tc-poor S stars - believed to be the cool descendants of barium stars. They find that the binary fraction, period distribution and eccentricities for all these classes differ from the typical orbital elements found among red giants in open clusters \citep[][for an updated sample]{Mermilliod96,Mermilliod07} in that they have a lower maximum eccentricity at a given period, which might be a sign of mass transfer. On the other hand, the orbital properties of the various barium and Tc-poor S star subsamples discussed in \citet{Jorissen98} are similar to those found by \citet{McClure90} and to each other.  

While it has
been hypothesized before that also CEMP-s and CH-stars as a class show a
common origin, the excellent agreement in their binary properties
shown here in this work puts this hypothesis on a much firmer footing. 

\begin{figure*}
\includegraphics[width=0.32\linewidth]{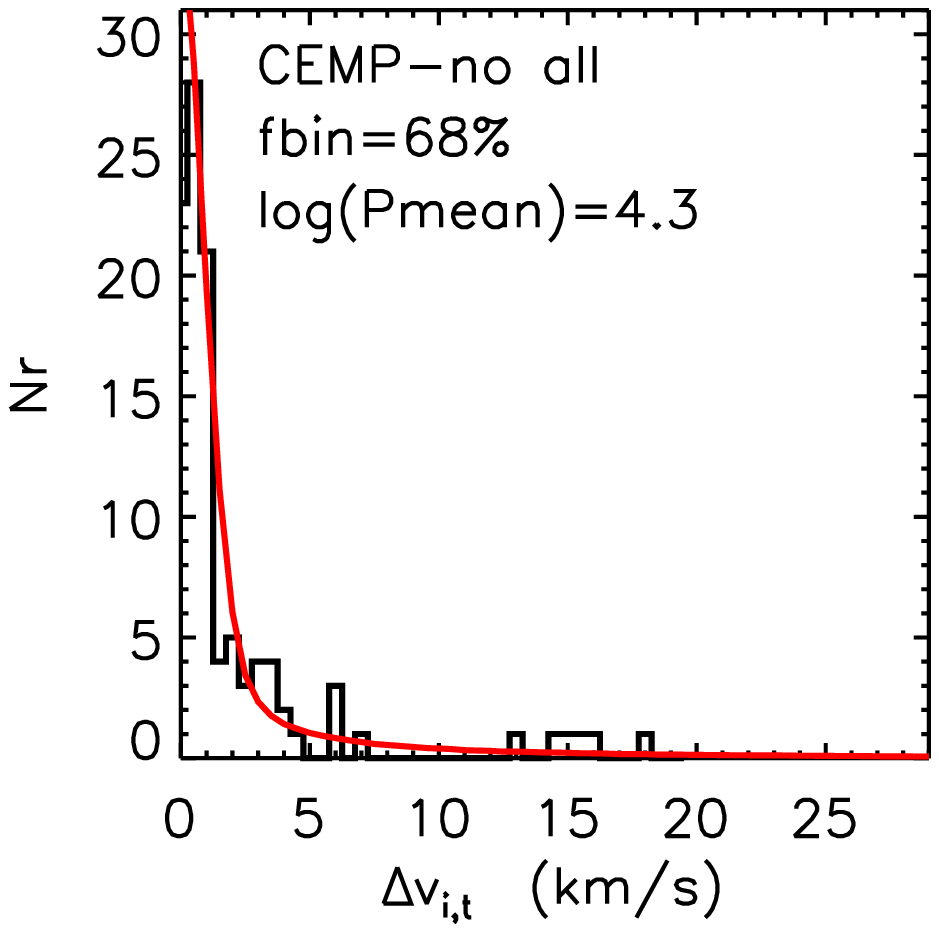}\includegraphics[width=0.32\linewidth]{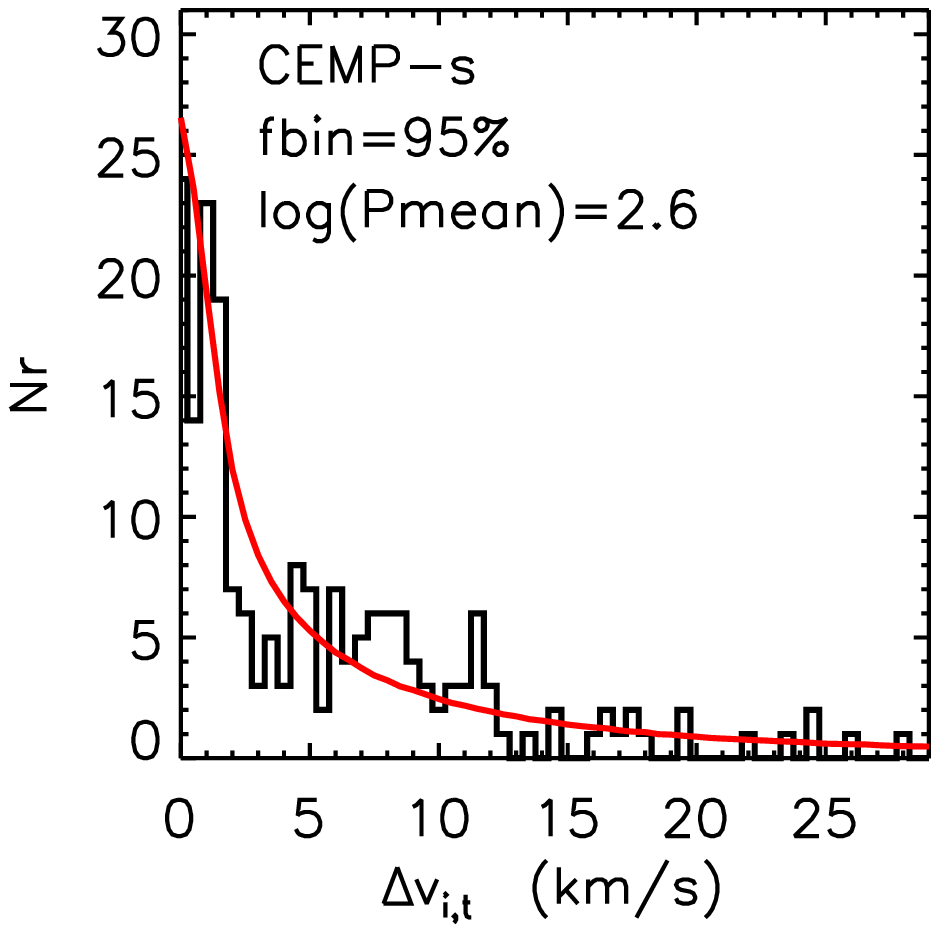} \includegraphics[width=0.32\linewidth]{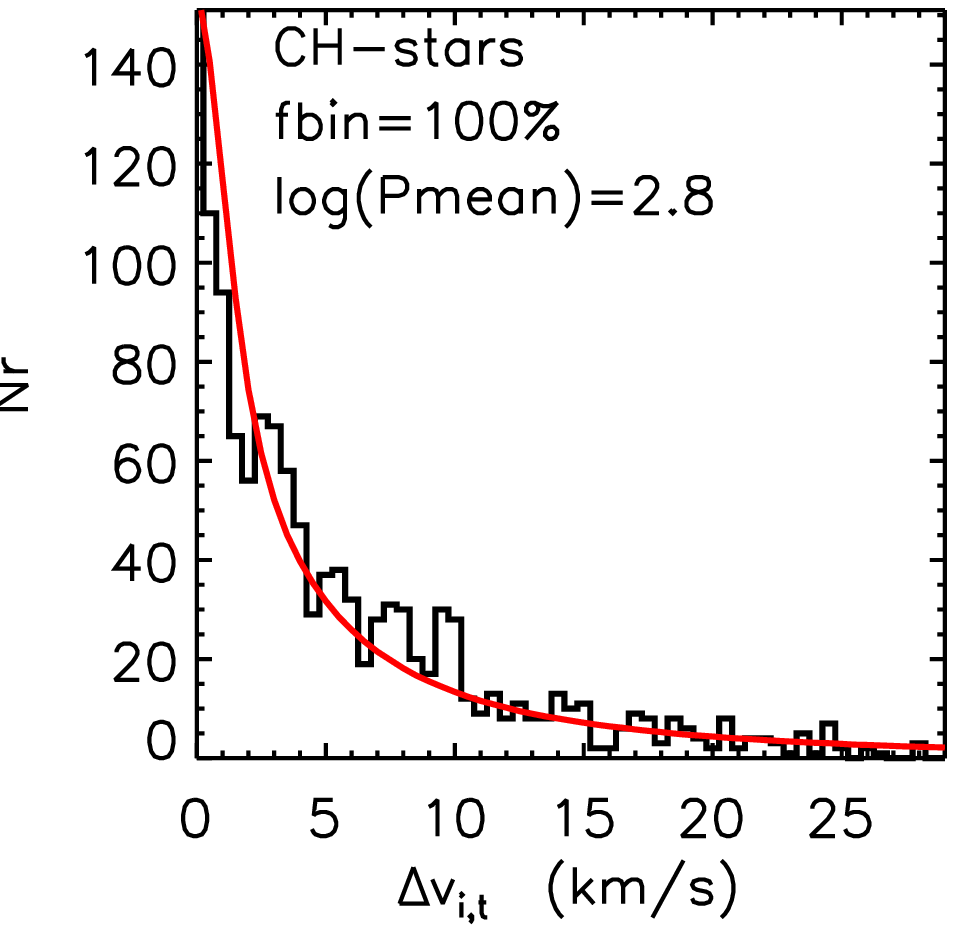} 
\caption{Histograms of each of distributions in $\Delta  v_{i,t}$ for
  the CEMP-no measurements from this work and combined and from the literature samples of CEMP-s, CH-stars,
  with overplotted the most probable distributions from the
  simulations. The best-fitting parameters for the binary fraction
  and mean period are labelled in the panels.\label{fig:hists}}
\end{figure*}

\subsection{All datasets compared}
Another key result from the analysis presented above is that, with very high
significance, the binary properties of
the CEMP-s and CH-stars, as a population, do not overlap with
that of the CEMP-no stars; as can be clearly seen when one compares
Figure \ref{fig:simcempno} with Figure \ref{fig:simcemps}.

Figure \ref{fig:hists} shows the histograms of $\Delta v_{i,t}$ for the CEMP-s, CH-stars, all CEMP-no and the CEMP-no stars in this work with overplotted their best solution according to the Bayesian posterior probability detailed above. These figures highlight again the differences in the distributions between these datasets in a more direct way, most particularly between the measurements of the CEMP-s and CH-stars on the one hand and the CEMP-no samples on the other.

\section{Discussion:What are the CEMP-no stars?}\label{sec:disc}

In this section we
will further investigate the remaining possible scenarios for the
origin of the peculiar chemical composition of the CEMP-no population.

\subsubsection*{They are born with it.}  
    The fact that some of the CEMP-no stars are in binaries does not automatically imply that mass
    transfer has to have happened. Perhaps the secondary has not
    evolved through an AGB phase, or mass transfer has not been
    effective. If we assume a similar distribution of periods,
    eccentricities etc., then -- according to the observational
    evidence as presented in Figure \ref{fig:simcempno} -- a somewhat
    higher binary fraction than in the Solar Neighbourhood is
    favoured. This would be in agreement with claims that binarity is
    generally higher in lower metallicity populations
    \citep[e.g.][]{Carney03}. So one could argue that the binarity of
    these stars are as expected from star formation processes in
    ``normal'' populations, and need not be connected to their peculiar
    chemical properties. Several scenarios in the literature have been
    suggested in which these stars form out of a birth cloud enhanced
    with carbon-enhanced material and deficient in s-process
    material. We refer the reader to \citet{Norris13b}, \citet{Karlsson13} and \citet{Nomoto13}
    for a detailed description of proposed scenarios and their
    (sometimes subtle) differences. Here, we instead highlight two
    main ideas. First, various models argue that the region could be
    enriched in C-rich material by the supernova and/or wind ejecta of
    Pop III stars where rotation of these stars plays a major role
    \citep[e.g.,][]{Fryer01,Meynet06,
      Chiappini06,Karlsson06,Hirschi07,Chiappini08,Meynet10,Maeder12,Cescutti13}. A
    second possibility involves a pristine gas cloud enriched by
    preceding supernovae explosions of more massive stars with
    fall-back, thereby locking in most heavy elements and expelling
    mainly lighter elements
    \citep{Umeda03,Limongi03,Iwamoto05,Umeda05,Tominaga07}. As
      presented in \citet{Ito13}, the detailed abundance pattern of
      the brightest CEMP-no star, BD +44-493, seems to be consistent
      with this scenario and not with pollution from an AGB companion or by massive, fast-rotating stars.
 
As discussed by \citet{Bromm03} once an enhancement in carbon and
    other light elements exists in a very metal-poor environment, this
    will aid the star formation process by shortening the timescale
    for cooling significantly \citep[see
    also][]{Frebel07b,Gilmore13}. 
The binary properties for CEMP-no
    stars are, in each of these scenarios, unrelated to their
    chemical history. We note that one prediction
    of this scenario is that should one be able to find and measure
    the properties of the binary companion; it will have a similar
    chemical composition as it most probably originated in the same
    birth cloud. Either of these scenarios
     suggest that the formation of CEMP-no stars is related to higher
     mass progenitors than of CEMP-s stars. As remarked by
     \citet{Carollo14} and \citet{Lee14}, studying the CEMP-no to CEMP-s ratio in
     various Galactic populations could therefore provide interesting clues
     about their Intitial Mass Function. On the other hand, differences in
     CEMP-no to CEMP-s stars in various populations could also be related to differences in
     binary fraction or properties, as we show in this work.

\subsubsection*{They are polluted.} 
From our analysis we find there is certainly some binarity present in the CEMP-no sample. Moreover, based on all radial velocity
    data currently in hand, we can not rule out that all CEMP-no
    stars are in binaries. However, if they are, many of these binary
    systems will have to be very wide. This questions the possibility
    of a scenario which involves mass transfer. We have convincingly shown that the
    binary population of CEMP-no stars is very different to that of
    the well-known classes of mass transfer binaries as CEMP-s and CH-stars. So, if CEMP-no stars would obtain their carbon
    excess from a companion, one would be tempted to look for a
    different transfer mechanism that would be able to act over a
    wider range of separations. As detailed in the previous section,
   it is somewhat debated if the mass transfer happens through
   Roche-lobe overflow, wind accretion or a combination. Each of the
    mechanisms proposed would have a typical range of separations, and
    therefore periods, over which it would be most effective. 

Additionally, we note that pollution processes do not necessarily involve binarity. Various of the pollution scenarios for the star's birth cloud as mentioned in the previous paragraph could also possibly pollute a star's atmosphere later on. One interesting candidate for such enrichment might be again massive stars going through a Wolf-Rayet (hereafter WR) phase, especially the sub-class of WR 
 carbon stars. While individual WR stars enrich only small local bubbles $<$ 10 pc in size
\citep[e.g., WR16,][]{Duronea13}, WR stars within star forming regions can
contribute to much larger bubbles.  The star forming region LMC N51D
\citep[or DEML192,][]{Davies76} contains UV-bright O-stars and a WC5
star \citep{Oey98}, and produces a wind-blown super-bubble
120 pc in size traced by several indicators including H$\alpha$ and
[S II] HST imaging \citep{Chen00}, XMM-Newton X-ray imaging
\citep{Bomans03}, and Spitzer IRAC imaging \citep{Chu05}.
Starburst galaxies with many concentrated WR stars can show
enrichments over even larger (kpc) scales. For instance, HST imaging
of NGC\,5253 by \citet{Calzetti04} find S[II] filaments $>1$ kpc
from the ionizing starburst at the centre.   While the C-enrichment
and stellar wind strengths of WR stars are found to be metallicity
dependent \citep[e.g.,][]{Portinari98, Maeder12},
the yields and stellar wind properties of massive rotating stars
are predicted to be effective at all metallicities.  Thus, whether from WR
stars themselves or from the cumulative effects of the stellar winds
from massive stars (especially massive rotating stars), it appears that carbon enrichments without neutron-capture
elements could occur over large scales in star forming regions over 
a range of metallicities.

\subsubsection*{During their evolution they accreted dust-depleted
  gas.} Another mechanism that could lead to a carbon enhancement without
neutron-capture elements would be the separation of gas and dust beyond
the stellar surface, e.g., the formation of a debris disk, followed by
the accretion of dust-depleted gas.
\citet{Venn08} compared the chemical abundances of two of
the metal-poor C-rich stars to those of the chemically peculiar
post-AGB, RV Tau, and Lambda Boo stars.  There are some similarities
that correlate with dust condensation temperature, and could imply
that grain formation contributes to the chemical abundance pattern,
rather than variations being due to natal or binary characteristics.
However, one critical test of this hypothesis is in the abundance
of sulphur and/or zinc, which have been found to be enriched in the
chemically peculiar stars, or more accurately, they have not been depleted
onto dust grains due to their low condensation temperatures. Unfortunately,
the upper limits on sulphur and zinc in most of the metal-poor C-rich stars
are insufficient to test this hypothesis; only one star, CS~22949-037
([Fe/H] = $-4.0$), has upper limits on these two elements that are
1.5 dex below the expected values \citep{Spite11}. For this star,
its metal-poor and C-rich nature cannot be explained by the separation of dust and gas in the stellar 
envelop. Also, [Zn/Fe] is low in BD +44-493 (Ito et al. 2013) suggesting that it has not seen dust formation either. A second critical test of this hypothesis is the presence of infrared or sub-mm emission from the dust 
grains. Spectral energy distribution fitting for several ultra metal-poor CEMP-no stars show no excess emission below 22 microns (Venn et al. 2014, in preparation), consistent with a lack of circumstellar material. These observations cannot rule out cooler debris disks though, such as those found around close-in A-stars and Lambda Bootis stars \citep[e.g., see][]{Wyatt08, Booth13}. In conclusion, the separation of dust and gas in a cool debris disk is
a valid hypothesis still for some - but definitely not all - CEMP-no stars.

\subsubsection*{They are the low-metallicity counterparts of R-stars.}
As the CH-, sgCH- and Ba II-stars are the higher
metallicity counterparts of CEMP-s (and possibly CEMP-r/s) stars, can
we point out a high-metallicity equivalent for CEMP-no stars? Such a
connection could help us to understand their formation mechanism. As
proposed also by \citet{Cohen06,Cohen13}, the
intriguing class of R-stars comes to mind (the ``R''-classification
hereby refers back to the R, N system of \citet{Cannon18} and should
not be confused with stars rich in r-process material). As described
in \citet{Dominy84}, these stars show carbon enhancements, but no
s-process enhancements. Additionally and curiously, they show no signs
\textit{at all} of binarity at a very high significance level
\citep{McClure97b}. As also shown in this work, the binary fraction
among CEMP-no is most certainly not zero, as for the R-stars. The
absence of binarity among R-stars led \citet{McClure97b} to argue that
perhaps their absolute absence of binarity -- much unlike any other
population of stars we know -- is a clue to their formation
history. He proposed these stars might have been in very close
binaries and that a merging event between the companions might have
led the carbon produced in the helium-core to be mixed outward. A
subsequent prediction from this scenario is that there will be no
sub-giant or dwarf R-stars (as they will have to be producing carbon in
order for it to be available for mixing). This prediction seems to
hold. This again implies that there is no connection between R-stars and
CEMP-no stars. Unlike the R-stars, CEMP-no stars seem to be distributed over much of the HR-diagram, including the main-sequence and sub-giant branch. 

\subsubsection*{They are a mixed bag.} 
The definition of the class of CEMP-no stars
    by \citet{Beers05} is mainly motivated by observational evidence, and might
    actually harbour stars from a variety of formation mechanisms that
    share a common carbon excess -- and a common barium depletion ([Ba/Fe]$<$0). As
    noted before, a star can belong to multiple classes depending on
    its ratio of abundances in Ba, Eu and Fe. The possibility of a combination of formation mechanisms in the CEMP-no class is intriguing when looking at the data set
from this work (right panel of Figure \ref{fig:simsustheir}). Here we identify some stars clearly as short-period
binaries, while the other stars are consistent with being single. We have searched
    the spectra of our close binary stars HE~1150-0428 and HE~1506-0113
    for any sign of double lines, but did not find any signatures,
    most likely indicating that their companion is fainter (which would be expected if the companion went through an AGB phase already and is now a white dwarf). Could
it be that in our CEMP-no class, truly
second-generation stars (born with a peculiar chemistry that includes
high carbon-enhancement) are mixed with stars that obtained their
carbon enhancements later in their lives, by a mass-transferring
binary or other mechanisms? As discussed before, mass-transfer mechanisms that would transfer carbon -- but no
or few s-process elements -- are found in several model predictions
for rotating or massive AGB stars \citep{Herwig03,Herwig04,Siess04}, or AGB-stars with very high neutron-to-Fe-peak-element seed ratios \citep{Busso99,Cohen06}.

It is worth pointing out here that the various stars in this CEMP-no sample, even though they
are all enhanced in carbon and depleted in [Ba/Fe], do not show
a similar chemical pattern for other elements. This might superficially be taken as
an indication for a variety of origins among these stars. Most theoretical models describing the origin of these stars focus on the explanation of the abundance pattern of one of the four hyper-metal-poor stars ([Fe/H]$<-$5), and do not simultaneously explain the full sample as researched here \citep[but see][who provide an individual best-fit Pop III SN model for 48 extremely metal-poor stars]{Tominaga14}. 

However, continuing the thought that there exist different formation mechanisms into one class, it must be remarked that we also see
no striking similarities in the patterns of those that appear to be in
close binaries. For instance, the two stars with outstanding radial velocity variations, HE~1150-0428 and
HE~1506-0113, show very different abundance patterns, even though they
have very similar metallicities and comparable temperatures and
gravities \citep{Yong13a,Cohen06}. HE~1506-0113 is only moderately
enhanced in [N/Fe], and shows significant enhancements in Na and Mg. On
the other hand, HE~1150-0428 shows a higher [C/Fe] and very high
enhancement in [N/Fe], solar-like [Na/Fe], but enhancements in [Ti/Fe]
and in particular [Ca/Fe] \citep[see also Figure 7. of][for a direct
comparison]{Norris13b}. Both stars are depleted in [Sr/Fe], but
HE~1150-0428 much less so. Unfortunately, neither of the two stars has
a robustly measured [Eu/Fe] abundance, making it difficult to comment
on the possibility that these might also be of the CEMP-r or CEMP-r/s
class. With the (lack of) present evidence it is very well possible that the CEMP-no class of stars includes any combination of the above mentioned formation scenarios. Among other avenues, we expect further insight in these mysterious stars to be obtained in the coming years by a careful analysis of their abundance patterns, focussing on the abundances of light elements as well as their neutron-capture elements \citep[see for instance the work of][emphasizing the importance of the lighter elements]{Masseron12,Norris13b, Ito13, Hansen14, Keller14}.

\section{Conclusions: The relationship between the various classes
  of carbon-stars}

In this work we have investigated the binary properties of three
sub-classes of carbon enhanced stars for which radial velocity
monitoring is available: The CEMP-s stars that are metal-poor and show
enhancements in carbon and barium; the combined group of CH-stars
(consisting of (sg)CH- and Ba II - stars) that share these abundance
signatures at higher metallicities; and the CEMP-no stars, metal-poor stars
with enhancements in carbon but not in barium. For the latter group we
have presented new radial velocity monitoring from the HET, thereby
greatly improving the available information on these mysterious
stars. With this new data two CEMP-no stars show clear variabilities in their radial
velocities, indicating they are part of binary system. For various
other stars more data is needed. 

From a comparison of the data available for each of these subgroups
with simulations in which the binary fraction and maximum period from
the period distribution is treated as a free parameter, we can draw
the following conclusions:

\begin{itemize}
\item{Binary properties in CEMP-no stars are marginally
    consistent with the observed
    Solar Neighbourhood binary fraction and periods.}
\item{\textit{If} CEMP-no stars are all in binaries, some of them have
    very long periods. A solution in which the binary fraction is
    lower, but the binaries have shorter periods, is also very
    likely.  }
\item{Binarity of CEMP-s stars is well-modeled with an (almost) 100\%
    binary fraction and a maximum
    period $\sim$20,000 days.}
\item{CEMP-s stars and CH-stars share similar binary
    properties. This places the hypothesis that CEMP-s stars are the
    lower metallicity equivalents of the CH-stars on a much firmer
    footing.}
\item{CEMP-no and CEMP-s stars have very different binary
    properties, therefore it is unlikely their overabundance in carbon
    is obtained via the same physical mechanism.}
\item{The CEMP-no population are not the
    metal-poor equivalents of R-stars. A primary origin for the carbon
    enhancement remains very likely, although an origin from a binary
    companion using a mechanism that can operate for long period
    systems and which does not transfer s-process elements
    can not yet be ruled out. Another distinct possibility is that the
  CEMP-no class contains various physical sub-classes in itself.}
\end{itemize}

\section*{Acknowledgments}
We thank Falk Herwig and David Hartwick for very helpful discussions and the referee for insightful suggestions that helped to improve the manuscript. The authors are indebted to the International Space Science Institute 
(ISSI), Bern, Switzerland, for supporting and funding the international 
team ``First stars in dwarf galaxies''. E.S. also gratefully
acknowledges the Canadian Institute for Advanced Research (CIFAR)
Global Scholar Academy and the Canadian Institute for Theoretical Astrophysics
(CITA) National Fellowship for partial support. The Hobby-Eberly Telescope (HET) is a joint project of the University of Texas at Austin, the Pennsylvania State University, Stanford University, Ludwig-Maximilians-Universit\"{a}t M\"{u}nchen, and Georg-August-Universit\"{a}t G\"{o}ttingen. The HET is named in honor of its principal benefactors, William P. Hobby and Robert E. Eberly.

\bibliography{references}

\end{document}